\documentclass[10pt]{article}
\usepackage{geometry}
\usepackage{cite}
\usepackage{amssymb,amsmath}
\usepackage{wrapfig}
\usepackage{graphicx}
\usepackage[small]{caption}
\usepackage[title]{appendix}
\usepackage{authblk}
\usepackage{abstract}
\newtheorem{lemma}{Lemma}

\begin{document} 
\title{On convertibility among bipartite 2x2 entangled states}
\author{Yiruo Lin \thanks{yiruolin@tamu.edu, linyiruo1@gmail.com}}
\affil{\textit{Department of Computer Science \& Engineering, Texas A\&M University}}
\date{}
\maketitle

\begin{abstract}
\normalsize
Some progress is reported on conditions for convertibility among bipartite 2x2 entangled states: An inconvertibility condition related to the rank of an entangled state is given that it is impossible to convert to an entangled state with lower rank under separable operations; a particular set of local operations and classical communication (LOCC) is used to analyze convertibility of three subclasses of states - Werner states, Bell diagonal states and maximally entangled mixed states (MEMS). It is conjectured that MEMS may lie on the bottom of entangled state ordering for given entanglement of formation. A plausible way is suggested of systematically calculating convertibility in a general subclass of bipartite states whose density matrices are defined to be diagonal in a common basis. The set of LOCC adopted in this work is argued to be generalizable to provide sufficient conditions for convertibility among a large range of general 2x2 entangled states.
\end{abstract}

\section{Introduction} \label{intro}

Knowledge of convertibility condition between a pair of entangled states of a bipartite quantum system under various sets of physical operations is of both theoretical and experimental interest. We may obtain a deeper physical understanding of entanglement characterizing a general entangled state by examining its convertibility to other entangled states under various operations on the state. As entanglement has been well appreciated as a useful resource in many quantum information and quantum computation tasks, knowledge of convertibility among entangled states under experimental accessible operations is desirable for ordering usefulness of various entangled states at hand. \\

Convertibility condition under local operations and classical communication (LOCC) between two multiple copies of bipartite entangled states allowing local operations on multiple systems is uniquely determined by entanglement entropy in the asymptotic limit\cite{asymp}. In practice, one always deals with a finite amount of copies and sometimes no local collective measurement is available which is the case in the following discussions. In general, there is no unique physical quantity providing complete information on the convertibility. A sufficient and necessary condition was obtained in \cite{Nielson} for convertibility between a pair of pure bipartite states under LOCC. Later work further expanded the result of \cite{Nielson} to obtain criteria for convertibility from a pure bipartite state to a mixed bipartite state and separable operations were found to be equivalent to LOCC (which is strictly enclosed by separable operations) in determining the convertibility condition. In \cite{sep_LOCC}, effect of separable operations compared to LOCC on convertibility from a mixed entangled state to a pure entangled state was examined. To the best of the author's knowledge, the condition for convertibility between two general bipartite entangled states is still unknown.  \\

In this work, we report progress and put forth some thoughts on convertibility among bipartite entangled states. We focus on convertibility among bipartite entangled states of dimension 2x2, which is still a hard problem to solve. We first generalize the arguments in \cite{sep_LOCC} to derive an inconvertibility condition among bipartite entangled states under separable operations. We then restrict our attention to several subclasses of entangled states and derive some results by making use of a particular set of LOCC. Finally we conjecture promising avenues of making further progress. The paper is organized as follows. In section \ref{inconvert}, we discuss a general inconvertibility condition among 2x2 bipartite entangled states. In section \ref {subclass}, we discuss convertibility conditions for three subclasses of 2x2 bipartite states. We conclude with discussions on future directions of work in section \ref {discussion}. \\ 

\section{An Inconvertibility Condition on 2x2 Entangled States} \label{inconvert}

We make use of the insight on the effect of local state dimensionality on convertibility from mixed to pure entangled states to extend the work in \cite{sep_LOCC} and derive a further condition under which certain conversion is forbidden under separable operations. For any bipartite entangled state, the reduced density matrix of each local system must have rank$>$1, namely, it corresponds to a mixed local state. Hence, for any separable operation on the entangled state which doesn't destroy its entanglement, at least one member of the complete set of general local positive valued measurements (POVM) needs to have rank$>$1 matrices on both local systems. When one of the local Hilbert space has dimension 2, this implies a full rank matrix on the corresponding local system. The full rank identity of the local operator matrix is what essentially leads to inconvertibility of any mixed 2x3 state to a pure entangled state\cite{sep_LOCC}.  \\

Making use of the role of restricted dimensionality of local Hilbert space, we derive a situation under which conversion is forbidden among general 2x2 entangled states. We state the result as:

\begin{lemma}
\label{mixed 2x2 states} It is impossible to convert under separable operations a 2x2 entangled state to another 2x2 entangled state if the latter has lower rank. 
\end{lemma}
The proof is given in the Appendix \ref{lemma 1}. This result generalizes the previous result given in \cite{sep_LOCC} which is a special case where it is impossible to convert a mixed 2x2 entangled state  (i.e., rank$>$1) to a pure entangled state (i.e., rank$=$1). It is interesting to observe that in contrast to the pure bipartite state conversion where it is impossible to convert a state with lower Schmidt rank to one with higher Schmidt rank, this lemma suggests that in converting among mixed entangled states, the von Neumann entropy of the state tends to increase and the purity tends to decrease. \\

\section{Convertibility conditions on three subclasses of bipartite 2x2 entangled states} \label{subclass}

For convertibility among general bipartite 2x2 states, a complete classification was known for pure to pure and pure to mixed states for which LOCC coincides with separable operations. The result of \cite{sep_LOCC} rules out any conversion from mixed to pure states. So it remains to see how to order 2x2 mixed entangled states and whether there is any difference under separable operations compared to LOCC. A general 2x2 mixed state is completely characterized by 9 independent parameters up to local unitary transformations \cite{mixed_state}. Hence the minimal number of a complete set of monotones (i.e., functions that are non-increasing under given operations) is no less than 9 \cite{common_cause} to fully characterize convertibility among 2x2 mixed states, which is much more complicated compared pure state and pure to mixed state conversion (for which one monotone suffices). Given the complexity of the problem, we will constrain ourselves to three subclasses of mixed states including Werner states \cite{Werner}, Bell diagonal states and maximally entangled mixed states (\textbf{MEMS}) \cite{MEMS_1}. They all have interesting physical properties and mathematically relatively simple to calculate.  \\

We make use of the same set of LOCC operations to address all of them. The set is made of a mixture of local unitary operations and operations that discard the given state and prepare a separable state. \\

\subsection{Werner states}

A Werner state of local dimension 2 is a mixture of a singlet state with identity matrix (completely mixed state) and can be parameterized by one parameter quantifying the degree of mixing. It is straightforward to show that any single valid entanglement monotone suffices to order Werner states since one can explicitly demonstrate the convertibility between any two Werner states by LOCC. A mixture of identity operation (i.e., do noting) and an operation that discards the given state followed by preparing completely mixed state (realizable by LOCC) suffices to convert a Werner state to another Werner state with lower weight of singlet state. It is obvious that separable operations can not induce more conversion than LOCC. \\

\subsection{Bell diagonal states}

The density matrix of a Bell diagonal state has all of its eigenvectors Bell states and can be parameterized, up to local unitary transformations, by three parameters, say the largest three eigenvalues. In \cite{SLOCC}, a complete set of monotones was found that guarantees probabilistic conversion from one Bell diagonal state to another when all of the monotones are non-increasing. Thanks to high local unitary symmetry of Bell diagonal states, any separable operation leading to a Bell diagonal state probabilistically from another Bell diagonal state can be realized by a mixture of a few extremal operations up to local unitary transformations. The extremal operations belong to two types: they are either local unitary transformations or projections onto a two-dimensional subspace spanned by two of the Bell states which is then discarded and replaced by a separable state (made of an equal mixture of the two projected Bell states). The local unitary transformations are obviously reversible and can be implemented with certainty. On the other hand, although projection operations are in general irreversible and probabilistic,  the subsequent operation (discard and preparation of a separable state) can be implemented with certainty. Therefore, we can nevertheless replace a probabilistic operation due to projection by a deterministic operation which discards the given state and replaces it with a separable state. All we need to do is to renormalize the weights of the original probabilistic operations for the new deterministic operations. Thus, we show that any probabilistic operation for transforming among Bell diagonal states can be realized by a deterministic operation. For example, consider a probabilistic operation consisting of equal mixture of a local unitary transformation and a projection followed by preparing a separable state. Suppose the projection succeeds with probability of 1/2. The overall probability of transformation for this probabilistic operation is 3/4. We can replace this operation by a deterministic one with renormalized weight of 2/3 for the local unitary transformation and 1/3 for the deterministic operation of preparing the separable state from scratch.  So we have at hand a complete set of monotones that can certify the transformation among Bell diagonal states deterministically under LOCC, not only probabilistically \cite{Note_Bell-diag}. We state the result as follows: 
\begin{lemma}
\label{Bell diag} Let $\rho$ and $\rho'$ be two entangled Bell diagonal states with, respectively, weight vectors $\vec{\lambda}=(\lambda_1,\lambda_2,\lambda_3, \lambda_4)^T$ and $\vec{\lambda'}=(\lambda'_1,\lambda'_2,\lambda'_3, \lambda'_4)^T$ in the nonascending order, i.e., $\rho=\sum_{i=1}^4\lambda_i\Pi_i$ and $\rho'=\sum_{i=1}^4\lambda'_i\Pi_i$, $\Pi_i $ are projectors onto the Bell basis. Conversion from $\rho$ to $\rho'$ via LOCC can be realized iff 
\begin{eqnarray}
E_1(\vec{\lambda})&\geq& E_1(\vec{\lambda'}), \nonumber \\
E_2(\vec{\lambda})&\geq&E_2(\vec{\lambda'}), \nonumber \\
E_3(\vec{\lambda})&\geq&E_3(\vec{\lambda'}), \label{Bell_diag}
\end{eqnarray}
where 
\begin{eqnarray}
E_1(\vec{\lambda})&\equiv& \lambda_1, \nonumber \\
E_2(\vec{\lambda})&\equiv&\frac{1-2\lambda_2}{\lambda_3+\lambda_4}, \nonumber \\
E_3(\vec{\lambda})&\equiv&\frac{1-2\lambda_2-2\lambda_3}{\lambda_4}. \label{def_E}
\end{eqnarray}
\end{lemma}
The three monotones $E_1(\vec{\lambda})$, $E_2(\vec{\lambda})$ and $E_3(\vec{\lambda})$ come from Theorem 4 of \cite{SLOCC}. The above analysis also shows that separable operations are no more powerful than LOCC since any probabilistic separable operation can be replaced by deterministic LOCC. \\

\subsection{MEMS}

Another interesting class of mixed states is MEMS, which can be written as a mixture of a singlet state with separable states parameterized by three parameters. Again, by LOCC made of a mixture of local unitary rotations and a local operation which discards the given state and prepares a separable state, we can convert a MEMS with a higher weight of singlet state to one with a lower weight. When the conversion is between rank 2 states,  a sufficient and necessary condition is easily obtained. Conversion among general MEMS satisfies an additional constraint. The result is stated as follows:
\begin{lemma}
\label{MEMS} Let $\rho$ and $\rho'$ be two MEMS with eigenvalues $\lambda_i$ and $\lambda'_i$, $i=1,\cdots,4$ in the non-ascending order respectively. When $\mathrm{rank}(\rho)=\mathrm{rank}(\rho')=2$, the sufficient and necessary condition for conversion from $\rho$ to $\rho'$ under LOCC is $\lambda_1>\lambda'_1$. When $\mathrm{rank}(\rho)=\mathrm{rank}(\rho')=3$, a sufficient condition for conversion from $\rho$ to $\rho'$ under LOCC is $\lambda_1-\lambda_3>\lambda'_1-\lambda'_3$ and $\frac{\lambda'_1-\lambda'_3}{\lambda_1-\lambda_3}=\frac{\lambda'_1}{\lambda_1}$. When $\mathrm{rank}(\rho')=4$, a sufficient condition for conversion from $\rho$ to $\rho'$ is given by $\lambda_1-\lambda_3>\lambda'_1-\lambda'_3$, $W\lambda_4+(1-W)(1-\eta_1-\eta_2)=\lambda'_4$ with $W=\frac{\lambda'_1-\lambda'_3}{\lambda_1-\lambda_3}$, $\eta_1=\frac{\lambda'_2-W\lambda_2}{1-W}$ and $\eta_2=\frac{\lambda'_3-W\lambda_3}{1-W}$.
\end{lemma}
The proof is given in the Appendix \ref{lemma 3}.

\section{Conclusion} \label{discussion}

Convertibility condition among general mixed entangled states is still an open problem. Even for the simplest 2x2 states, one needs at least 9 entanglement monotones to order them. We reported here some progress in this direction. We obtained an inconvertibility condition on general 2x2 entangled states and convertibility conditions for three subclasses of 2x2 states. The inconvertibility condition suggests a trend of increasing von Neumann entropy and decreasing purity of a state during conversion. On the other hand, MEMS has the interesting property that it maximizes von Neumann entropy and minimizes purity for a given entanglement of formation\cite{MEMS_2}. This in turn suggests MEMS may lie on the bottom of entangled state ordering given the same entanglement of formation. To further develop this conjecture, it would be interesting to start with examining conversion between Bell diagonal states and MEMS fixing the entanglement of formation. \\

Using a simple set of LOCC, we obtained the sufficient and necessary condition for convertibility among Werner states, among Bell diagonal states and among rank 2 MEMS. The set of LOCC can only provide a sufficient but not necessary condition for convertibility among general MEMS. To make further progress on MEMS, the methodology adopted in \cite{SLOCC} may provide a viable approach. Roughly speaking, by Choi–Jamiołkowski isomorphism, the separable density matrix corresponding to a given separable operation takes a diagonal form for both the target bipartite system and the auxiliary bipartite system (cf. equation (3) and Figure 1 of \cite{SLOCC}) when we only consider conversion among a subset of 2x2 states written in the diagonal space (i.e. the subset is defined to be diagonal in this space). Since the set of all separable matrices forms a polytope, we need to find its vertices in terms of which a general separable density matrix can be cast. A sufficient and necessary convertibility condition may then be calculated. In general, the condition is only valid for a probabilistic conversion. To ensure a deterministic conversion, trace preserving condition needs to be applied to the separable density matrix. It would be interesting to check whether the observation is just a coincidence that a probabilistic conversion can always be replaced by a deterministic one in the case of Bell diagonal states. The methodology looks promising to provide a systematic way of ordering subsets of 2x2 states with three independent parameters. \\

The set of LOCC used in categorizing the three subclasses of 2x2 entangled states may provide useful information on the convertibility between two general 2x2 entangled states. By \cite{BSA}, any 2x2 state can be optimally decomposed into a mixture of a single pure entangled state and a separable state. If the entangled state is the same (up to local unitary rotations) in the decomposition for the two states in consideration, then the set of LOCC can be used to give a sufficient convertibility condition. In particular, a necessary condition for the decomposed pure entangled state to be a  maximally entangled state is given by Lemma 5 of \cite{BSA1}. If the separable state in a decomposition has full rank, then the state can always be decomposed into a mixture of a pure maximally entangled state and a separable state. In such a case, we may apply the set of LOCC to provide a sufficient condition for convertibility. \\

\section*{Acknowledgement}

The author thanks the support from the Department of Computer Science \& Engineering, Texas A\&M University.


\begin{appendices}
\section{Proof of Lemma 1} \label{lemma 1}
Consider two 2x2 entangled states $\rho$ and $\rho'$ shared by Alice and Bob. Suppose under a separable operation $\sum_iE_i(\cdot)E_i^\dagger$ with $E_i=A_i\otimes B_i$ and $\sum_iE^\dagger_iE_i=I$, $\rho$ is converted to $\rho'$. We now prove that the conversion is impossible if $\mathrm{rank}(\rho)>\mathrm{rank}(\rho')$. The case of a pure $\rho'$ is already proved, we only need to prove $\mathrm{rank}(\rho')=2, 3$. When $\mathrm{rank}(\rho')=2$, consider without loss of generality $\mathrm{rank}(\rho)=3$. Since the operation converts to an entangled state, there exists $A\otimes B$ with both A and B full rank such that 
\begin{eqnarray} 
A\otimes B|\psi_1\rangle &\propto& |\phi_1\rangle, \nonumber \\
A\otimes B|\psi_2\rangle &\propto& a_1|\phi_1\rangle+b_1|\phi_2\rangle, \nonumber \\
A\otimes B|\psi_3\rangle &\propto& a_2|\phi_1\rangle+b_2|\phi_2\rangle, \label{rank3to2}
\end{eqnarray}
where $|\psi_i\rangle$, $i=1,2,3$ is the eigenvectors of $\rho$, $|\phi_1\rangle$ and $\phi_2\rangle$ are orthogonal to each other.  From equation (\ref{rank3to2}), we have $I\otimes B(\sum_{i=1}^3c_i|\psi_i\rangle)=0$. But this is impossible since $\sum_{i=1}^3c_i|\psi_i\rangle\neq0$ and $B$ is of full rank. The case of $\mathrm{rank}(\rho')=3$ can be similarly proved.

\section{Proof of Lemma 3} \label{lemma 3}

Consider two MEMS $\rho=(\lambda_1-\lambda_3)|\psi_s\rangle\langle\psi_s|+\lambda_3(|00\rangle\langle00|+|11\rangle\langle11|)+\lambda_2|01\rangle\langle01|+\lambda_4|10\rangle\langle10|$ and $\rho'=(\lambda'_1-\lambda'_3)|\psi_s\rangle\langle\psi_s|+\lambda'_3(|00\rangle\langle00|+|11\rangle\langle11|)+\lambda'_2|01\rangle\langle01|+\lambda'_4|10\rangle\langle10|$, where $|\psi_s\rangle$ is a singlet state, $\lambda_i$ and $\lambda'_i$, $i=1,\cdots,4$ are ordered nonascendingly. \\

When $\mathrm{rank}(\rho)=\mathrm{rank}(\rho')=2$, i.e., $\rho=\lambda_1|\psi_s\rangle\langle\psi_s|+(1-\lambda_1)|01\rangle\langle01|$ and $\rho'=\lambda'_1|\psi_s\rangle\langle\psi_s|+(1-\lambda'_1)|01\rangle\langle01|$, if $\lambda_1>\lambda'_1$, we can convert $\rho$ to $\rho'$ by the LOCC that mixes an identity operation with a operation that discards the given state and prepares $|01\rangle$. To prove that it is also necessary, note that the entanglement of formation is given by $\lambda_1$ and $\lambda'_1$ respectively\cite{MEMS_max}. \\

When $\mathrm{rank}(\rho)=\mathrm{rank}(\rho')=3$, apply LOCC to $\rho$ with weight W of identity operation and weight 1-W of an operation that discards the given state and prepares a separable state of form $\eta|01\rangle\langle01|+(1-\eta)(|00\rangle\langle00|+|11\rangle\langle11|)$, require the resulting state be $\rho'$, we obtain
\begin{eqnarray}
W(\lambda_1-\lambda_3)&=&\lambda'_1-\lambda'_3, \nonumber \\
W\lambda_3+(1-W)(1-\eta)&=&\lambda'_3, \nonumber \\
W\lambda_2+(1-W)\eta&=&\lambda'_2, \label{rank3_MEMS}
\end{eqnarray}
$\frac{\lambda'_1-\lambda'_3}{\lambda_1-\lambda_3}=\frac{\lambda'_1}{\lambda_1}$ follows. \\

When $\mathrm{rank}(\rho')=4$, a similar consideration as above is applied and we obtain
\begin{eqnarray}
W(\lambda_1-\lambda_3)&=&\lambda'_1-\lambda'_3, \nonumber \\
W\lambda_3+(1-W)\eta_2&=&\lambda'_3, \nonumber \\
W\lambda_2+(1-W)\eta_1&=&\lambda'_2, \nonumber \\
W\lambda_4+(1-W)(1-\eta_1-\eta_2)&=&\lambda'_4. \label{rank4_MEMS}
\end{eqnarray}
\end{appendices}

\end{document}